\pdfoutput=1
\documentclass[aps,prd,print,preprintnumbers,showpacs,superscriptaddress,twocolumn]{revtex4}
\usepackage{latexsym,amssymb,amsmath,mathrsfs}
\usepackage[pdftex]{graphicx}
\usepackage{setspace,bm,color}
\usepackage{hyperref}
\usepackage[caption=false]{subfig}
\graphicspath{{./Figures/}}

\newcommand{\diff}{\mathrm{d}}
\newcommand{\p}{\partial}
\newcommand{\ve}{\varepsilon}

\newcommand{\be}{\begin{equation}}      
\newcommand{\ee}{\end{equation}}      
\newcommand{\bea}{\begin{eqnarray}}      
\newcommand{\eea}{\end{eqnarray}}

\newcommand{\ka}{K{\" a}hler }
\newcommand{\im}{\mathrm{i}}

\begin{document}

\preprint{RIKEN-QHP-166}

\title{\boldmath Lefschetz-thimble techniques for path integral of zero-dimensional $O(n)$ sigma models}

\author{Yuya Tanizaki}
\email{yuya.tanizaki@riken.jp}
\affiliation{Department of Physics, The University of Tokyo, Tokyo 113-0033, Japan}
\affiliation{Theoretical Research Division, Nishina Center, RIKEN, Wako, Saitama 351-0198, Japan}

\date{\today}

\begin{abstract}
Zero-dimensional $O(n)$-symmetric sigma models are studied using Picard--Lefschetz integration method in the presence of small symmetry-breaking perturbations. 
Because of approximate symmetry, downward flows turn out to show significant structures: They slowly travel along the set of pseudoclassical points, and branch into other directions so as to span middle-dimensional integration cycles. 
We propose an efficient way to find such slow motions for computing Lefschetz thimbles. 
In the limit of symmetry restoration, we figure out that only special combinations of Lefschetz thimbles can survive as convergent integration cycles: Other integrations become divergent due to noncompactness of the complexified group of symmetry. 
We also compute downward flows of $O(2)$-symmetric fermionic systems, and confirm that all of these properties are true also with fermions. 
\end{abstract}


\maketitle

\section{Introduction}\label{sec:intro}
Path integral with Picard--Lefschetz theory \cite{pham1983vanishing,Witten:2010cx, Witten:2010zr} is now becoming popular and developing as a useful method for nonperturbative studies of quantum systems: It was first introduced in order to study analytic properties of gauge theories in coupling constants~\cite{Witten:2010cx, Witten:2010zr, Harlow:2011ny}. 
This new formulation of the path integral is also considered to be useful for studying quantum tunneling phenomena directly using the real-time path integral~\cite{Tanizaki:2014xba, Cherman:2014sba}, establishing resurgence trans-series methods of perturbation theories~\cite{Unsal:2012zj, Basar:2013eka, Cherman:2014ofa, Dorigoni:2014hea}, 
and solving the sign problem of Monte Carlo simulations of lattice field theories~\cite{Cristoforetti:2012su, Cristoforetti:2013wha, Cristoforetti:2014gsa, Aarts:2013fpa, Fujii:2013sra, Nishimura:2014rxa, Mukherjee:2014hsa, Aarts:2014nxa}. 

Symmetry is a universal clue to studying quantum systems, and symmetry breaking is an especially important aspect of quantum statistical physics. 
So far, the continuous symmetry breaking has not yet been addressed from the perspective of Picard--Lefschetz integration methods. 
Therefore, studying symmetries and their breaking using this new approach must broaden future applicabilities of this formalism to various physical systems.  

In the Picard--Lefschetz integration method, integration cycles are deformed into ``nicer'' ones, called Lefschetz thimbles, in a complexified space of the original integration cycle. This procedure makes the convergence of integration much better, which therefore opens a new way to evaluate the path integral. 
Since it relies on complex analogue of Morse theory, critical points of the classical action must be nondegenerate. 
If a continuous symmetry exists, however, classical solutions are degenerate due to the symmetry, and a special treatment is required. 
In Ref.~\cite{Witten:2010cx}, a general method for a system with unbroken symmetries is already established, which will be briefly reviewed in Sec.~\ref{sec:form} for a specific example. 

Our primary purpose in this paper is to consider the effect of small symmetry-breaking perturbations on Picard--Lefschetz integration. 
In Sec.~\ref{sec:symm_break_o2}, we consider a zero-dimensional $O(2)$-symmetric sigma model, and propose an efficient method to compute Lefschetz thimbles with a small symmetry-breaking term. 
Because of approximate symmetry, some of the downward flows turn out to slowly travel along the set of approximate critical points, and we show that this is an important ingredient for Lefschetz thimbles in the perturbed system. 
We scrutinize these properties of the flow equation, and figure out the global behavior of Lefschetz thimbles in our model. 
Furthermore, by comparing Lefschetz thimbles in symmetric and nonsymmetric systems, we can obtain a better understanding of Lefschetz thimbles for symmetric systems. 
In Sec.~\ref{sec:symm_break_on}, generalization of our analysis for $O(n)$ symmetries is discussed, and those analyses are reduced into those for $O(2)$ symmetry. 

In Sec.~\ref{sec:fermion}, we consider another zero-dimensional sigma model including fermionic variables, which looks like the BCS theory of superfluidity. 
After integrating out fermions, the effective bosonic action has logarithmic singularities, and this model shows the sign problem. 
We confirm all the properties of downward flows associated with symmetry, and thus our proposal to compute Lefschetz thimbles turns out to work well also with fermions. 

This paper is organized as follows. In Sec.~\ref{sec:form}, we give a basic setup of the model, and Lefschetz thimbles with continuous symmetry are briefly reviewed. 
In Sec.~\ref{sec:symm_break}, the properties of Lefschetz thimbles are studied in detail for the zero-dimensional bosonic model when the continuous $O(n)$ symmetry is slightly broken explicitly. 
We confirm that these properties are true even with fermions in Sec.~\ref{sec:fermion}. 
Section~\ref{sec:summary} is devoted to a summary. 

\section{Lefschetz thimbles with symmetry}\label{sec:form}
In this section, we review applications of Picard--Lefschetz theory to oscillatory integrations with unbroken continuous symmetries \cite{Witten:2010cx, Witten:2010zr}. 
For simplicity, we concretely consider a zero-dimensional $O(n)$ sigma model. 

Let $\bm{\sigma}=(\sigma_a)_a\in \mathbb{R}^n$, and then consider the Lie-group action $O(n)\curvearrowright \mathbb{R}^n$ defined by the left multiplication. Its infinitesimal transformation can be characterized by real anti-symmetric matrices:  
\be
\delta_{\ve} \sigma_a =\ve_{ab}\sigma_b
\label{eq:gp_ac01}
\ee
with $\ve_{a b}=-\ve_{b a}\in\mathbb{R}$. 
We consider an action function $S_0:\mathbb{R}^n\to \mathbb{R}$, which is invariant under $O(n)$, and study the property of integration, 
\be
Z_0(\hbar)=\int_{\mathbb{R}^n}\diff^n \sigma~\exp(-S_0[\bm{\sigma}]/\hbar), 
\label{eq:form01}
\ee 
where $\diff^n\sigma=\diff \sigma_1\wedge\cdots \wedge \diff \sigma_n$ is the Lebesgue measure. 

Before trying Lefschetz-thimble methods to evaluate (\ref{eq:form01}), let us consider the result of $O(n)$ symmetry in a direct way. 
The $O(n)$ symmetry of the integrand and integration measure says that we can integrate radial and angular directions separately. 
By integrating out the angular directions, we just need to evaluate the following integration 
\be
Z_0(\hbar)={2\pi^{n/2}\over \Gamma(n/2)}\int_{0}^{\infty}\diff \sigma \,\sigma^{n-1}\exp(-S_0(\sigma)/\hbar), 
\label{eq:form02}
\ee
with $S_0(\sigma)=S_0[\sigma\hat{\bm{e}}]$ ($\hat{\bm{e}}$ is an arbitrary unit vector), instead of the original one (\ref{eq:form01}). 
We can interpret this result from the viewpoint of the reduction of integration cycles: 
The quotient space $(\mathbb{R}^n\setminus\{0\})/O(n)$ is identified with the positive real axis $\mathbb{R}_{>0}$.

In the rest of this section, we review a general way to derive (\ref{eq:form02}) without taking quotients of the original integration cycle \cite{Witten:2010cx}. 
This also provides a basic setup for taking into account the effect of explicit breaking terms. 

We take the complexification $\mathbb{C}^n$ of $\mathbb{R}^n$, and introduce the complex coordinate $\xi_a=\sigma_a+\im \eta_a$. The complexified space $\mathbb{C}^n$ canonically admits the symplectic structure~\footnote{In order for Picard--Lefschetz integration method, the complexified space must be chosen to admit a \ka metric, which is a closed, non-degenerate two-form and compatible with the complex structure. The important point is that the \ka metric uniquely provides the symplectic form $\omega$ in (\ref{eq:form03}). This gives the Poisson bracket and allows us to consider the Hamilton mechanics. }, 
\be
\omega=-{\im\over 2} \diff \xi_a\wedge \diff \overline{\xi_a}=\diff \eta_a \wedge \diff \sigma_a\,. 
\label{eq:form03}
\ee
We exploit the fact that a gradient flow for the Lefschetz thimble can be viewed as a Hamiltonian flow of classical mechanics \cite{Witten:2010cx} (see also \cite[Appendix]{Tanizaki:2014xba} for a review). 
Namely, we regard $\{\sigma_a\}_a$ as canonical coordinates and $\{\eta_a\}_a$ as conjugate momenta, and consider the Poisson bracket $\{f,g\}={\p f\over \p \sigma_a}{\p g\over \p \eta_a}-{\p f \over \p \eta_a}{\p g\over \p \sigma_a}$. 
Let us consider the Hamilton mechanics with the Hamiltonian 
\be
H_0=\mathrm{Im}\;S_0[\bm{\xi}],
\label{eq:form06}
\ee
then its equation of motion with respect to ``time'' $t$, ${\diff \xi_a\over \diff t}=\{H_0,\xi_a\}$, is nothing but Morse's downward flow equation, 
\be
{\diff \xi_a\over \diff t}=\overline{\left({\p \over \p \xi_a}S_0[\bm{\xi}]\right)}. 
\label{eq:form05}
\ee
Along the flow, the Hamiltonian $H_0=\mathrm{Im}\;S_0[\xi]$ as well as other constants of motion are conserved, which plays a pivotal role 
in the following discussions.  

The group action (\ref{eq:gp_ac01}) must also be extended. The real orthogonal group $O(n)$ is complexified to the complex orthogonal group $O(n,\mathbb{C})=\{A\in GL(n,\mathbb{C})| A A^T=\mathrm{id}_n\}$. 
The complexified group action $O(n,\mathbb{C})\curvearrowright \mathbb{C}^n$ is given by 
\be
\delta_{\widetilde{\ve}} \xi_a=\widetilde{\ve}_{ab}\xi_b, 
\label{eq:gp_ac02}
\ee
with $\widetilde{\ve}_{a b}=-\widetilde{\ve}_{b a}\in\mathbb{C}$. This extension makes clear that the $O(n)$ symmetry of the original action $S[\bm{\sigma}]$ is extended to the $O(n,\mathbb{C})$ symmetry of $S[\bm{\xi}]$. 
Under this $O(n,\mathbb{C})$ transformation, the symplectic structure (\ref{eq:form03}) transforms as 
\bea
\delta_{\widetilde{\ve}}\omega=\mathrm{Im}(\widetilde{\ve}_{ab})\;\diff \xi_a\wedge\diff{\overline{\xi_b}}, 
\label{eq:form07}
\eea
and thus it is invariant if and only if the infinitesimal parameter $\widetilde{\ve}_{ab}$ is real. 
Therefore, the symmetry of this Hamilton system is $O(n)$, although the Hamiltonian $H_0$ is invariant under $O(n,\mathbb{C})$. 
Noether charges of the $O(n)$ symmetry are 
\be
J_{ab}(\xi,\overline{\xi})=-\eta_a \sigma_b+\eta_b \sigma_a={\im\over 2}(\xi_a\overline{\xi_b}-\xi_b\overline{\xi_a}),  
\label{eq:form08}
\ee
and a collection of these Noether charges is called the momentum map, denoted by $\mu_{O(n)}$~\cite[Appendix~5]{arnold_mechanics}. 
It is important to notice that $\mu_{O(n)}=0$ on the original integration cycle $\mathbb{R}^n$. 
Since Lefschetz thimbles contributing to the original integration must be connected with $\mathbb{R}^n$ via a flow \cite{Witten:2010cx}, such Lefschetz thimbles must be connected to the region $\mu_{O(n)}=0$ through $O(n,\mathbb{C})$ transformations. 

In this paper, let us consider the simplest toy model: 
\be
S_0[\bm{\xi}]={e^{\im \alpha}\over 4}(\bm{\xi}^2-1)^2,  
\label{eq:form09}
\ee
with $\bm{\xi}^2=\xi_1^2+\cdots+\xi_n^2$. 
Here, $\alpha$ should be formally regarded as an infinitesimally small positive constant, $\alpha=0^+$, which makes Lefschetz thimbles well-defined as integration cycles. 
Critical points of $S_0[\bm{\xi}]$ consist of two sets: One of them is the origin $O=\{\bm{\xi}=0\}$, and the other one is a complex quadric,  
\be
Q^{n-1}=\{\bm{\xi}\in\mathbb{C}^n\;|\;\bm{\xi}^2=1\}. 
\label{eq:form10}
\ee
Complexified critical orbits can be nicely parametrized with slow variables. {For each symmetry, there exists two degenerate directions around a critical point: One of them is given by the symmetry transformation, and the another one by the corresponding Noether charge. This provides the easiest way to find the set of zero modes, or slow variables, of the flow equation (\ref{eq:form05}). } Mathematically, this implies the fact that $Q^{n-1}$ can be identified with the cotangent bundle $T^*S^{n-1}$ of the hypersphere $S^{n-1}=\{\bm{\sigma}\in\mathbb{R}^n\;|\; \bm{\sigma}^2=1\}$. 
We will see this explicitly in Sec.~\ref{sec:symm_break} for $n=2$. 

\begin{figure}[t]
\centering
\includegraphics[scale=0.3]{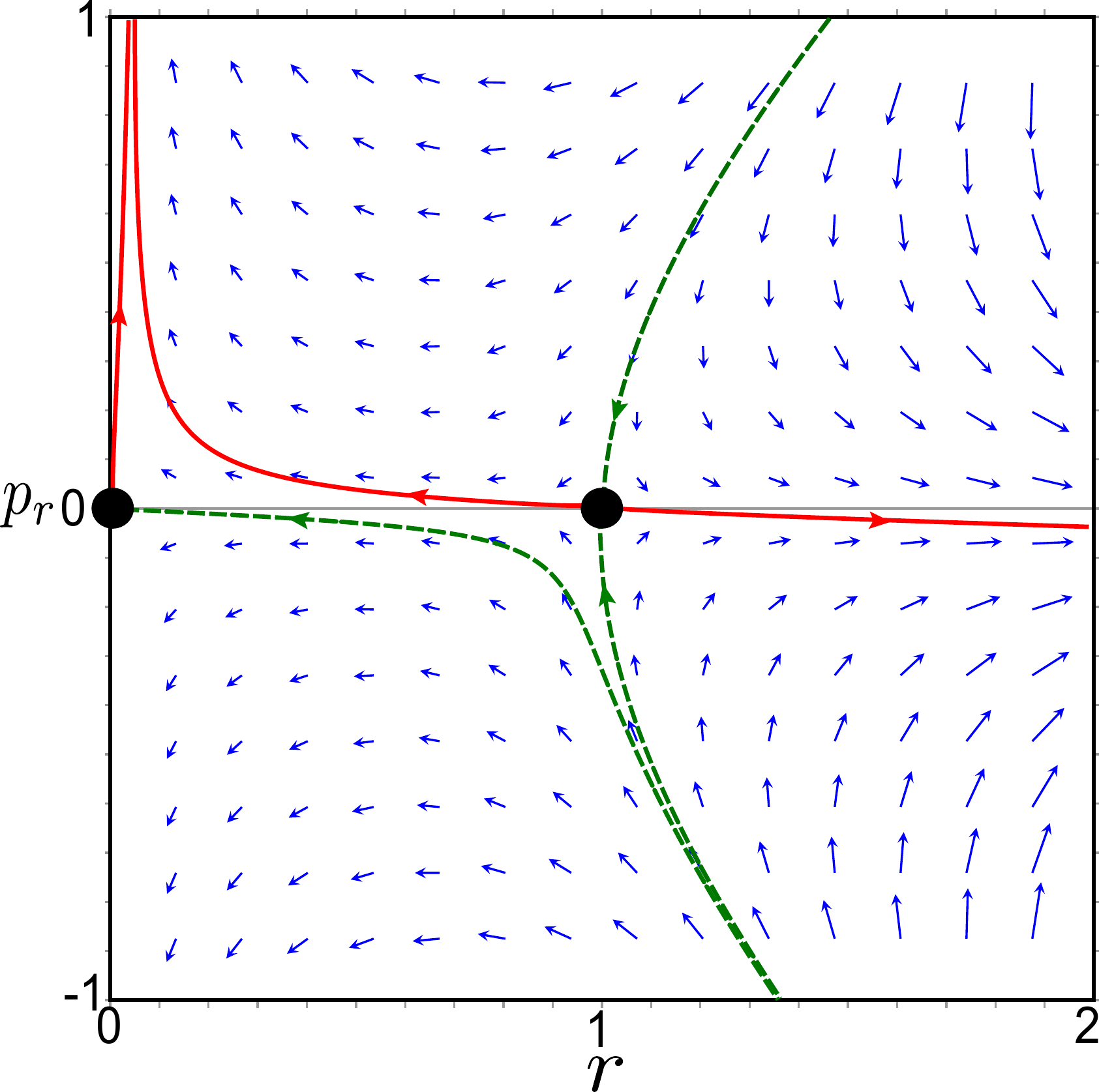}
\caption{Downward flows of the reduced Hamilton system ($p_{\theta}=0$) in the $rp_r$ plane with $\alpha=0.1$. Two points with $(r,p_r)=(0,0)$ and $(1,0)$ are critical points of this system. 
Red solid lines represent Lefschetz thimbles $\mathcal{J}$, i.e., downward flows emanating from critical points. Green dashed lines are their homological duals $\mathcal{K}$, which are characterized by downward flows getting sucked into critical points. 
Blue arrows show Hamiltonian vector fields.  
}
\label{fig:rprflow}
\end{figure}

Since the critical point $\xi=0$ is nondegenerate, we can compute its Lefschetz thimble in a usual way. 
Thus, we only explain how to construct the Lefschetz thimble of $\bm{\xi}^2=1$ in the following. 
Let us introduce the polar coordinate to emphasize the rotational symmetry by using the (point) canonical transformation: $r=|\bm{\sigma}|$ is a radial coordinate, $p_r=(\bm{\sigma}\cdot\bm{\eta})/|\bm{\sigma}|$ is its conjugate momentum, and $p_{\theta}^2=\sum_{a<b}J_{ab}^2$ represents the total angular momentum. 
The $O(n,\mathbb{C})$-invariant variable $\bm{\xi}^2$ can be written using $r$, $p_r$, and $p_{\theta}$ as 
\bea
\bm{\xi}^2&=&(\bm{\sigma}+\im\bm{\eta})^2=(\bm{\sigma}^2-\bm{\eta}^2)+2 \im \bm{\sigma}\cdot \bm{\eta}\nonumber\\
&=&r^2-\left({p_r^2}+{p_{\theta}^2\over r^2}\right)+2\im r p_r. 
\eea
The Hamiltonian $H_0$ of this $O(n)$-invariant system becomes  
\bea
H_0&=&{\sin \alpha \over 4}\left\{\left((r^2-1)- \left(p_r^2+{p_{\theta}^2\over r^2}\right )\right)^2 - 4r^2 p_r^2\right\}\nonumber\\
&&+\cos \alpha \;r p_r \left((r^2-1)- \left(p_r^2+{p_{\theta}^2\over r^2}\right)\right). 
\label{eq:form13}
\eea 
Because of the conservation law of $\mu_{O(n)}$, dynamics of the $rp_r$ direction is completely determined once $p_{\theta}^2$ is fixed. 
The result for the case $p_{\theta}^2=0$ is shown in Fig.~\ref{fig:rprflow}, and this is nothing but the downward flow for the integral~(\ref{eq:form02}) by regarding $r+\im p_r$ as a complexified variable of $\sigma$: $H_0|_{p_{\theta}=0}=\mathrm{Im}(S_0(r+\im p_r))$ with $S_0(\sigma)$ in (\ref{eq:form02}). 
The red and green dashed lines in Fig.~\ref{fig:rprflow} show downward flows emanating from and getting sucked into a critical point at $r+\im p_r=1$, respectively, which are given by 
\be
H_0(r,p_r)|_{p_{\theta}=0}=H_0(r=1,p_r=0)|_{p_{\theta}=0}. 
\ee
Once the radial motion is totally determined in this way, the angular motion is automatically determined, and thus we can embed Fig.~\ref{fig:rprflow} into $\mathbb{C}^n$. 
In order to obtain an $n$-dimensional convergent integration cycle $\mathcal{J}_{r=1}^{\mathrm{sym}.}$ and its dual $\mathcal{K}^{\mathrm{sym}.}_{r=1}$, we must pick up other $(n-1)$ directions from the $2(n-1)$-dimensional critical orbit $Q^{n-1}(=T^*S^{n-1})$. 

The convenient procedure to construct the Lefschetz thimble, which is proposed in \cite{Witten:2010cx}, is to rotate a red solid line of Fig.~\ref{fig:rprflow} by using $O(n)$ symmetry. 
As a result, the Lefschetz thimble $\mathcal{J}^{\mathrm{sym.}}_{r=1}$ for $r=1$ can be determined as the direct product of $S^{n-1}=\{\bm{\sigma}^2=1\}$ and the red solid line at $r=1$ of Fig.~\ref{fig:rprflow}: $\mathcal{J}^{\mathrm{sym}.}_{r=1}\simeq \mathbb{R}\times S^{n-1}$ as manifolds.  
Its dual $\mathcal{K}_{r=1}^{\mathrm{sym}.}$ is given by rotating the green dashed line at $r=1$ in Fig.~\ref{fig:rprflow} using the imaginary direction of $O(n,\mathbb{C})$. This procedure ensures that $\mathcal{J}$ and $\mathcal{K}$ intersect only at a single point. 
The reader may suspect that this construction contains some ambiguities because we pick up $(n-1)$ directions for $\mathcal{J}^{\mathrm{sym}.}_{r=1}$ by hand, but we can argue that this is the unique choice up to continuous deformations \cite{Witten:2010cx}. 
In Sec.~\ref{sec:symm_break}, we will see that this special choice of directions in $T^*S^{n-1}$ is consistent with the downward flows with symmetry-breaking perturbations. 

\section{Explicit symmetry breaking}\label{sec:symm_break}
In this section, we add a small symmetry-breaking term $\ve\Delta S$, which lifts degeneracies of critical points and allows us to use the familiar formulation of Picard--Lefschetz theory. 
Global behaviors of downward flows are scrutinized, and use of approximate symmetries plays an essential role for computation of Lefschetz thimbles. 
This study can become important in studying spontaneous symmetry breaking using Lefschetz thimbles.

\subsection{\boldmath $O(2)$ sigma model}\label{sec:symm_break_o2}

We again consider the $O(2)$ linear sigma model with the action (\ref{eq:form09}), and add a symmetry-breaking term, 
\be
\ve\Delta S[\bm{\sigma}]=\ve e^{\im \alpha} \sigma_1, 
\label{eq:break01}
\ee
with $0<\ve\ll 1$. We formally set $\alpha=0^+$ to simplify our argument. 
We study properties of the path integral 
\be
Z_{\ve}(\hbar)=\int_{\mathbb{R}^2} \diff \sigma_1\diff \sigma_2 \exp\left[-\left(S_0[\bm{\sigma}]+\ve\Delta S[\bm{\sigma}]\right)/\hbar\right], 
\label{eq:break02}
\ee
using Lefschetz-thimble techniques. 
We also discuss the relation to the previous formulation in the limit $\ve\to +0$ at the end of this section. 

Since there is an approximate $O(2)$ symmetry, the canonical transformation to the polar coordinate is again useful: 
\bea
  \begin{split}
    &r=\sqrt{\sigma_1^2+\sigma_2^2}\,,\; &&\theta=\tan^{-1}{\sigma_2\over \sigma_1}\,,
    \\
    &p_r={\sigma_1\eta_1+\sigma_2\eta_2\over \sqrt{\sigma_1^2+\sigma_2^2}}\,,\; 
    &&p_{\theta}=\eta_2\sigma_1-\eta_1\sigma_2\,. 
  \end{split}
  \label{eq:form11}
\eea
Here, $p_{\theta}$ is nothing but the Noether charge $J_{12}$ in (\ref{eq:form08}), and $\omega=\diff p_r\wedge \diff r+\diff p_{\theta}\wedge \diff \theta$. 
Since $\bm{\xi}^2=r^2+2\im r p_r-(p_r^2+p_{\theta}^2/r^2)$, the critical condition $\bm{\xi}^2=1$ can be solved as $p_r=0$ and $r^2-(p_{\theta}/r)^2=1$. Now, the $O(2,\mathbb{C})$ critical orbit (\ref{eq:form10}) is explicitly represented as 
\be
  \left\{
	\bm{\xi}=
    \left(\begin{array}{c}r\cos\theta-\im {p_{\theta}\over r}\sin\theta\\ r\sin\theta+\im {p_{\theta}\over r}\cos\theta\end{array}\right)
    \left|\begin{array}{c}r=\sqrt{ 1+\sqrt{1+4p_{\theta}^2} \over2}\,,\\ p_{\theta}\in\mathbb{R},\theta\in[0,2\pi)\end{array}\right.
  \right\}. 
  \label{eq:form12}
\ee
As we mentioned, this has a nice parametrization via slow variables $\theta$ and $p_{\theta}$, which implies that $Q^1\simeq T^* S^1$.   

The Hamiltonian of this system is $H=H_0+\ve \Delta H=\mathrm{Im}(S_0[\bm{\xi}]+\ve\Delta S[\bm{\xi}])$. 
The unperturbed Hamiltonian is given in (\ref{eq:form13}), and the perturbation is given by 
\be
\Delta H= \sin\alpha\; r \cos \theta+\cos\alpha\left(p_r \cos\theta-{p_{\theta}\over r}{\sin\theta}\right). 
\label{eq:break03}
\ee
This solves degeneracies of critical orbits, and validates Morse theory. However, naive computation of flows can easily fail, because of remnant of symmetry. 
After giving a proposal for circumventing this technical difficulty, we show its origin by scrutinizing properties of downward flows through concrete computations. 

\begin{figure}[t]
\centering
\includegraphics[scale=0.3]{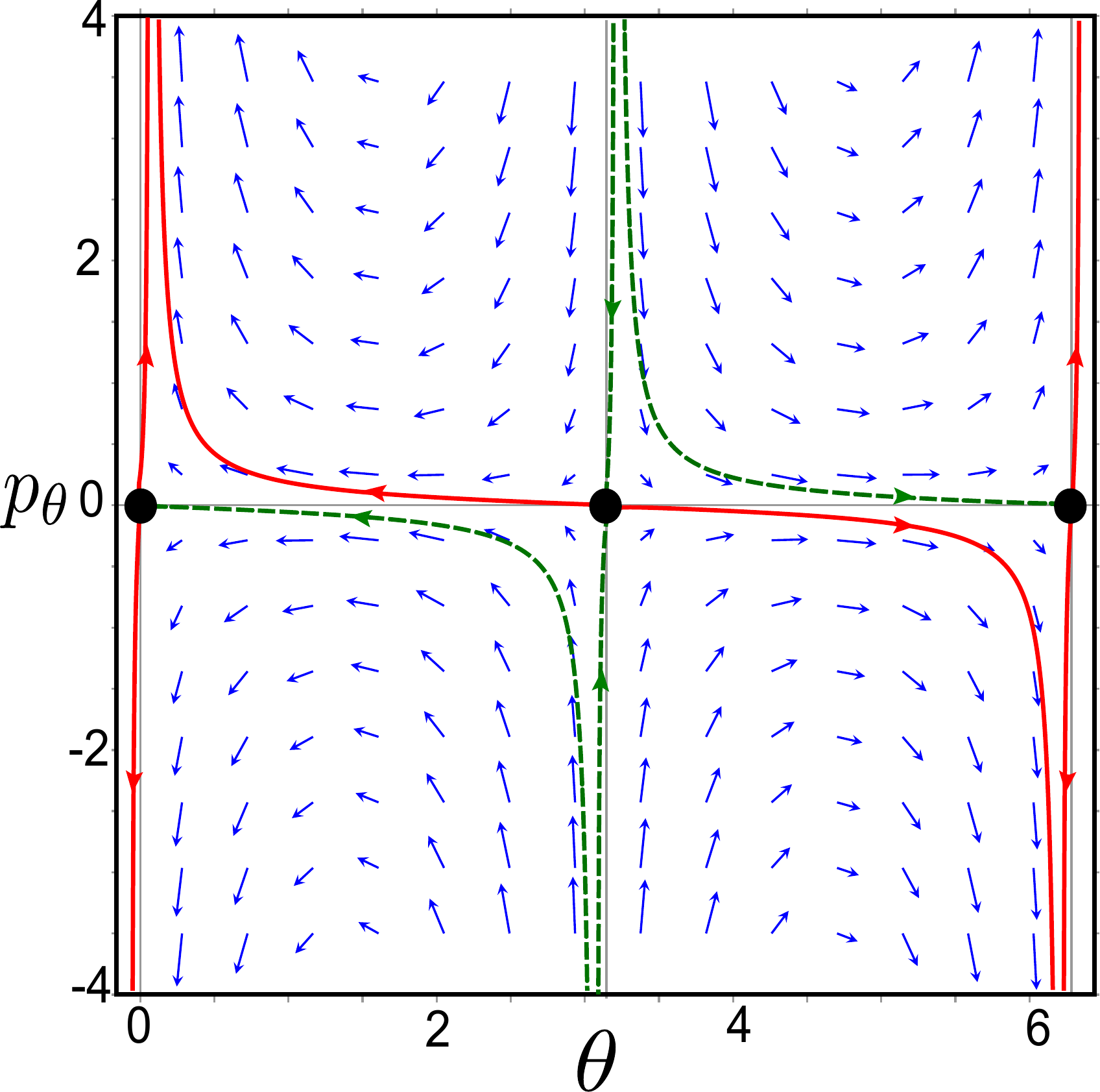}
\caption{Flows projected on the critical orbit $T^*S^1$ induced by the perturbation $\ve\Delta H$ at $\alpha=0.1$. Black circles at $\theta=0,\pi$ correspond to critical points in the perturbed system. Red solid and green dashed curves are downward flows projected onto $T^*S^1$, which emanate from and get sucked into critical points, respectively. Blue arrows show Hamiltonian vector fields. 
}
\label{fig:flow_critical_orbit}
\end{figure}

Before starting explicit computations, we mention an efficient way to compute Lefschetz thimbles and comment on behaviors of downward flows. 
We advocate that the downward flow on the pseudocritical orbit $T^* S^1$ should be calculated at first: 
The perturbation $\Delta H$ breaks the $O(2,\mathbb{C})$ symmetry of $S[\bm{\xi}]$, and it induces slow downward flows. 
The slow downward flow can be projected onto $T^*S^1$, which is shown in Fig.~\ref{fig:flow_critical_orbit}. 
In order to get Fig.~\ref{fig:flow_critical_orbit}, the Hamilton mechanics in terms of $\theta$ and $p_{\theta}$ is solved under the critical condition $p_r=0$ and $r=\sqrt{(1+\sqrt{1+4p_{\theta}^2\,}\,)/2}$. 
This step turns out to be important because Lefschetz thimbles in this system are two-dimensional cycles with one pseudoflat direction: 
To create such integration cycles, downward flows must first travel slowly along red solid curves on $T^*S^1$ in Fig.~\ref{fig:flow_critical_orbit} [up to $\mathcal{O}(\ve)$]. 
After that, flows branch into radial directions, as will be confirmed below. 

The critical condition $\p S/\p \xi_a=0$ gives 
\be
(\xi_1^2+\xi_2^2-1)\left(\begin{array}{c}\xi_1\\\xi_2\end{array}\right)+\left(\begin{array}{c}\ve\\0\end{array}\right)=\left(\begin{array}{c}0\\0\end{array}\right). 
\ee
Critical points of this system consist only of three points: $(\xi_1,\xi_2)=(\ve,0)$, $(\pm 1-\ve/2,0)$ up to $\mathcal{O}(\ve)$, and two of them $(\pm 1-\ve/2,0)$ correspond to black blobs in Fig.~\ref{fig:flow_critical_orbit} ($\alpha=0.1$ in this figure). 
In the polar coordinate (\ref{eq:form11}), these solutions are $(r,\theta)=(\ve,0)$, $(1-\ve/2,0)$, and $(1+\ve/2,\pi)$ with $p_r=p_{\theta}=0$. Among these critical points, $\mathrm{Re}(S_0+\ve \Delta S)$ takes the smallest value at $(r,\theta)=(1+\ve/2,\pi)$, and the largest value at $(r,\theta)=(\ve,0)$. Because of the symmetry-breaking term, the value at $(r,\theta)=(1-\ve/2,0)$ is lifted up from the smallest one roughly by $2\ve\cos\alpha$($= 2\ve$). 

The Hamiltonian $H$ takes $0$ and $\pm\sin\alpha$ at $(r,\theta)=(\ve,0)$, $(1-\ve/2,0)$, and $(1+\ve/2,\pi)$, respectively, up to $\mathcal{O}(\ve^0)$, and thus flows cannot connect distinct critical points. 
At $\alpha=0^+$, however, these differences are infinitesimally small: The Lefschetz thimble $\mathcal{J}_{(1+\ve/2,\pi)}$ around $(r,\theta)=(1+\ve/2,\pi)$ can pass infinitesimally close to other critical points. 

The set of Hamilton equations (exactly at $\alpha=0$) is 
\begin{subequations}\label{Eq:coupled}
\begin{align}
{\diff r\over \diff t}&=r\left(r^2-{p_{\theta}^2\over r^2}-1\right)-3r p_r^2+\ve \cos \theta, 
\label{eq:hamilton01}\\
{\diff p_r\over \diff t}&=(1-3r^2) p_r+p_r^3-{p_r p_{\theta}^2\over r^2}-\ve {p_{\theta}\over r^2}\sin\theta,
\label{eq:hamilton02}\\
{\diff \theta\over \diff t}&=-{2p_r p_{\theta}\over r}-\ve{\sin\theta\over r}, 
\label{eq:hamilton03}\\
{\diff p_{\theta}\over \diff t}&=\ve\left(p_r \sin\theta+{p_\theta\over r}\cos\theta\right). 
\label{eq:hamilton04}
\end{align}
\end{subequations}
Flow equations at generic $\alpha$ are given in Appendix~\ref{app:flow_generic}. 

\begin{figure}[t]
\centering
\includegraphics[scale=0.3]{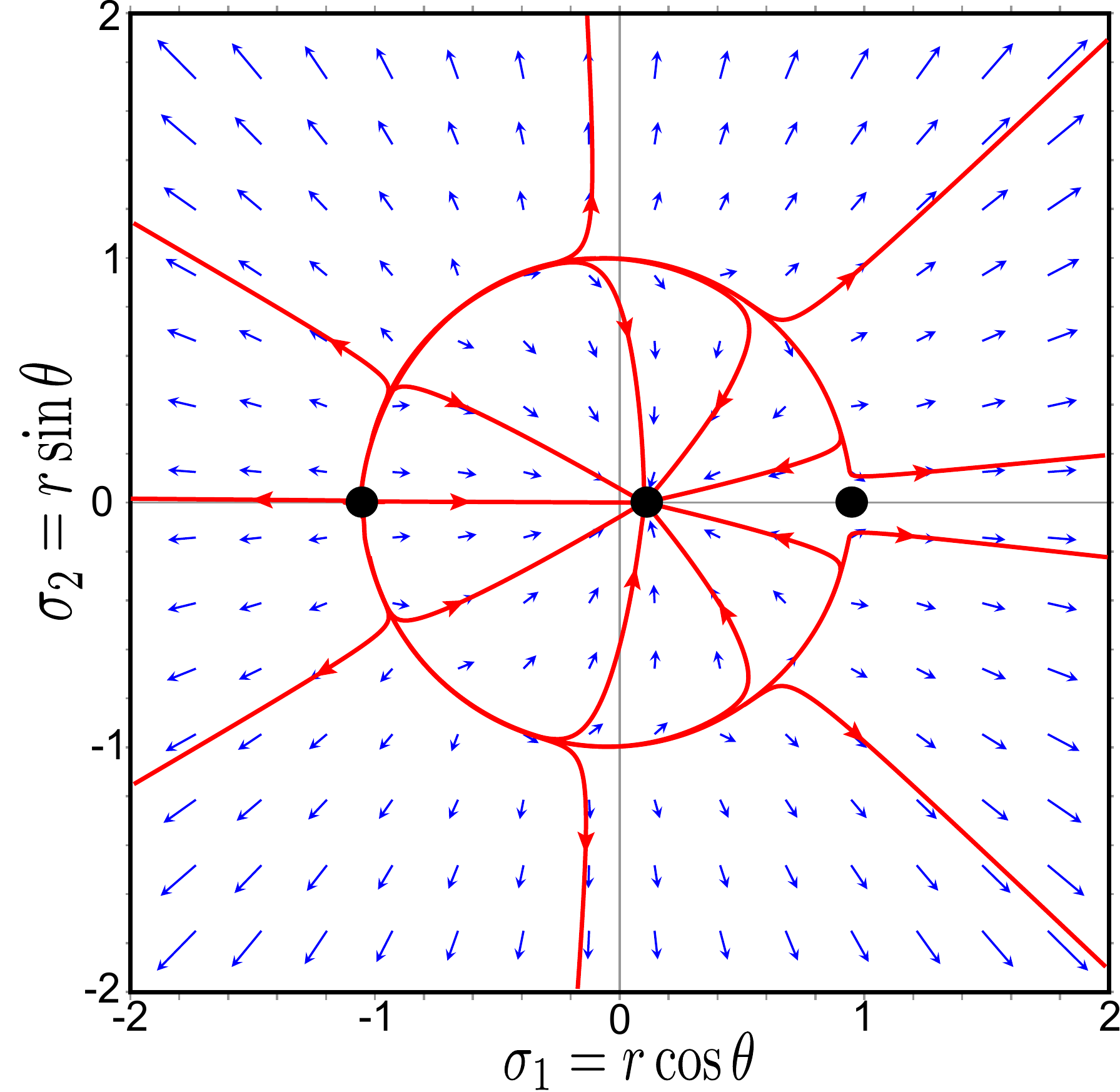}
\caption{Behaviors of downward flows (\ref{Eq:coupled}) in the $\sigma_1\sigma_2$-plane when $p_r=p_{\theta}=0$ and $\alpha=0$. 
Black blobs are critical points of the symmetry-broken system. 
Red arrowed curves are typical solutions starting from the global minimum $(r,\theta)=(1+\ve/2,\pi)$ with $0<\ve\ll 1$ ($\ve=0.1$ in this figure). 
Before branching into the $r$ direction, every flow from the global minimum moves slowly along the critical orbit $r=1$. 
}
\label{fig:xyflow}
\end{figure}

One can find a set of special solutions of this Hamiltonian system such that $p_r\equiv0$ and $p_{\theta}\equiv0$, which indeed automatically solves (\ref{eq:hamilton02}) and (\ref{eq:hamilton04}). 
Moreover, Figs.~\ref{fig:rprflow} and \ref{fig:flow_critical_orbit} imply that these are nothing but the conditions for a part of $\mathcal{J}_{(1+\ve/2,\pi)}$. 
Solutions of (\ref{Eq:coupled}) are shown in Fig.~\ref{fig:xyflow} in the original coordinate system $(\sigma_1,\sigma_2)=r(\cos\theta,\sin\theta)$. Let us consider flows starting from $(r,\theta)=(1+\ve/2,\pi)$. 
Near the critical point, one solution flows along radial direction in the same way as in the symmetric case, but there is another solution along $\theta$ direction. Indeed, setting $r\simeq 1$ in (\ref{eq:hamilton03}) gives 
\be
\theta(t)=\pm 2\tan^{-1}(\exp(-\ve t)). 
\label{eq:theta_motion}
\ee
According to (\ref{eq:theta_motion}), there is a downward flow rotating along a pseudoflat direction with a time scale $1/\ve$. For consistency with (\ref{eq:hamilton01}), the radial direction should be fine-tuned so that  $r(t)\simeq 1-(\ve/2)\tanh(\ve t) $. 
After traveling along the critical orbit $T^*S^1$ in this way, flows branch into radial directions, as shown in Fig.~\ref{fig:xyflow}: 

Drawing flow lines starting from a critical point is thus a difficult task, because we must designate the way of branching after traveling the pseudocritical orbit just by tuning the initial condition. 
Since we already know the approximate flow on the critical orbit in Fig.~\ref{fig:flow_critical_orbit}, however, we can start and revert a flow from a point of branch in drawing Fig.~\ref{fig:xyflow}. 
This procedure does not require fine-tuning of initial conditions. 
All we have to confirm is that the reverted flow is drawn into the critical point.

\begin{figure}[t]
\centering
\includegraphics[scale=0.3]{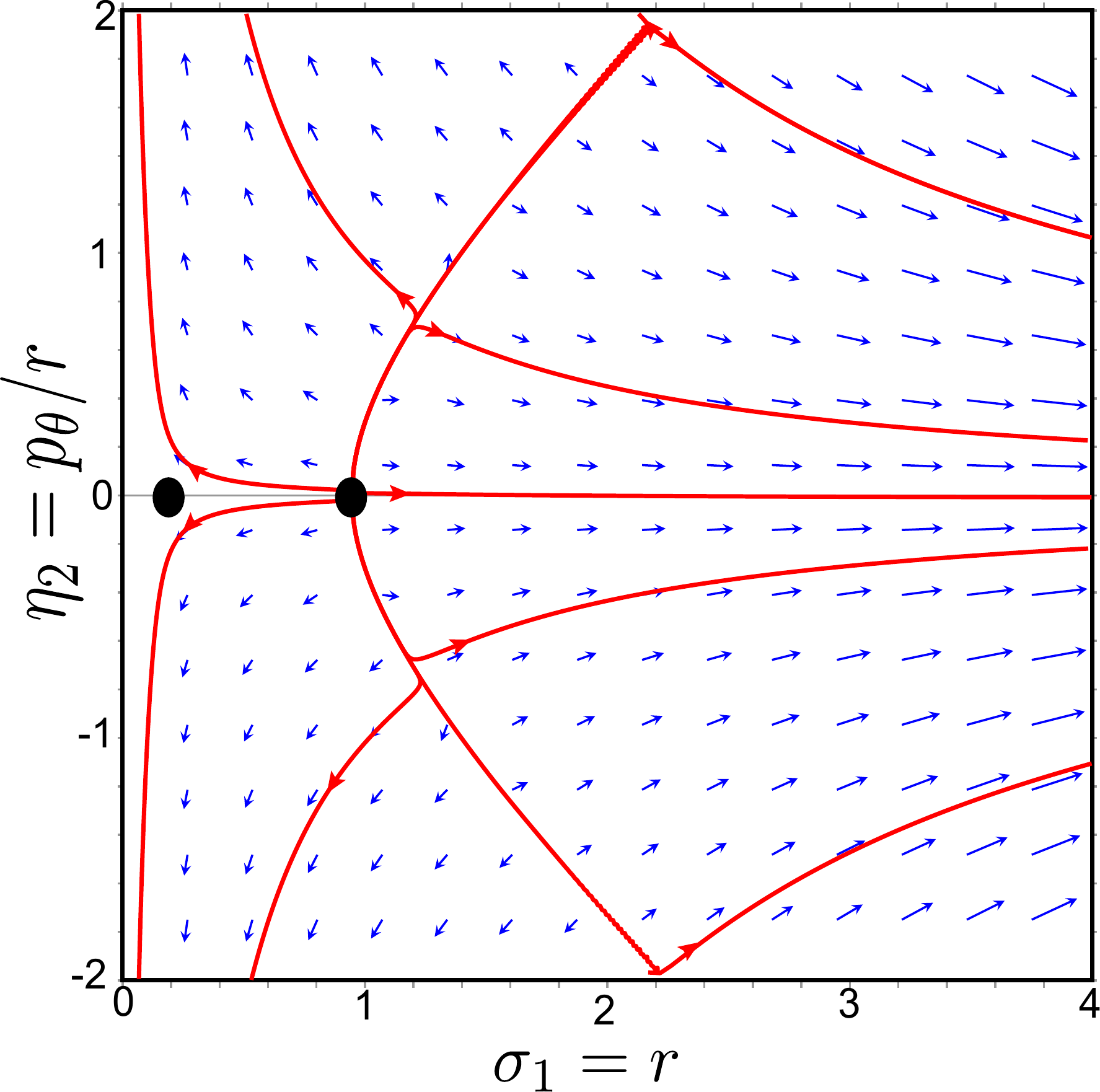}
\caption{Behaviors of solutions to the differential equation (\ref{Eq:coupled}) in the $\sigma_1\eta_2$-plane with $\theta=p_r=0$ when $\alpha=0$. Red arrowed curves show typical solutions starting from $(r,\theta)=(1-\ve/2,0)$ with $\ve=0.1$. }
\label{fig:xpyflow}
\end{figure}

Next, let us calculate the Lefschetz thimble $\mathcal{J}_{(1-\ve/2,0)}$ for $(r,\theta)=(1-\ve/2,0)$. 
For that purpose, we consider another set of special solutions of (\ref{Eq:coupled}) by putting $\theta=0$ and $p_r=0$, which gives the red solid line around $\theta=0$ in Fig.~\ref{fig:flow_critical_orbit} (but at $\alpha=0$). 
This condition solves (\ref{eq:hamilton02}) and (\ref{eq:hamilton03}) automatically. 
Behaviors of the downward flow in this plane are shown in Fig.~\ref{fig:xpyflow}, and,  
as we expected, downward flows go along the hyperbola $\sigma_1^2-\eta_2^2=1$ at first, which is characterized by $p_{\theta}$: Put $r(p_{\theta})=\sqrt{1+\sqrt{1+4p_{\theta}^2}\over 2}$, then (\ref{eq:hamilton04}) gives
\be
2r(p_{\theta}(t))-\tanh^{-1}(1/r(p_{\theta}(t)))=\ve t. 
\ee
After this slow motion, they branch into the $r$ direction in order to span a two-dimensional cycle $\mathcal{J}_{(1-\ve/2,0)}$. 
In drawing red flow lines in Fig.~\ref{fig:xpyflow}, we start and revert the flow from a point of branch instead of a point near the critical point. 
It would be really hard to control global behaviors of the flow if we compute the flow starting from a point near the critical point. 

\begin{figure}[t]
\includegraphics[scale=0.55]{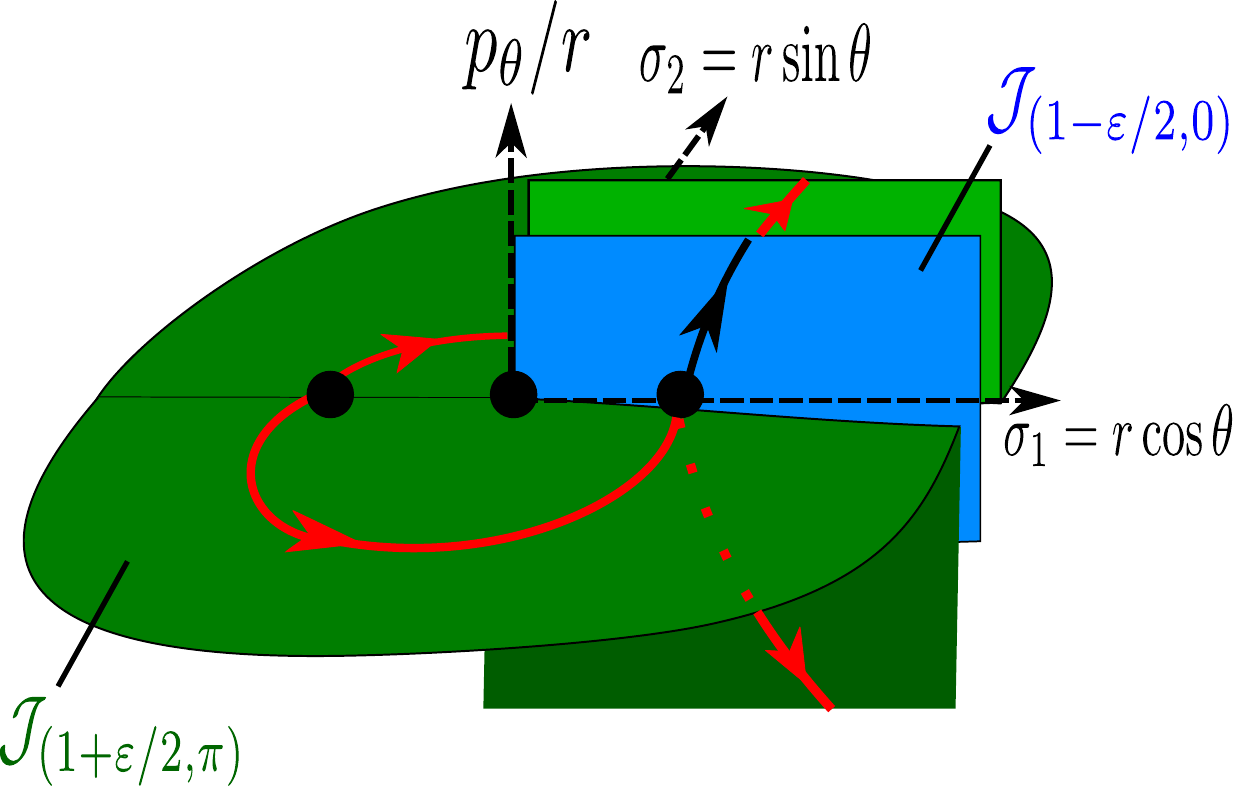}
\caption{Schematic figure for global structure of Lefschetz thimbles $\mathcal{J}_{(1+\ve/2,\pi)}$ and $\mathcal{J}_{(1-\ve/2,\pi)}$ at $\alpha=0^+$, which are shown with green and blue surfaces, respectively. 
In this limit, flows into $p_r$ direction of these thimbles are quite small unless $r\simeq 0$, and we totally neglect that direction. 
Three black blobs represent critical points of this system, and arrowed lines show slow motion with time scale $1/\ve$ along the critical orbit $T^*S^1$. 
}
\label{fig:schematic}
\end{figure}

Now that we have identified local behaviors of Lefschetz thimbles around critical points and properties of approximate downward flows on the pseudocritical orbit, we can comment on global behaviors of Lefschetz thimbles. This enables us to comment also on the relation of Lefschetz thimbles in symmetric and nonsymmetric cases. 
The Lefschetz thimble $\mathcal{J}_{(1+\ve/2,\pi)}$ around the global minimum goes away to infinities in $\eta_2$ direction  at $\theta=0,2\pi$ 
because there is another Lefschetz thimble $\mathcal{J}_{(1-\ve/2,0)}$ at $\theta=0$. See Fig.~\ref{fig:schematic} for its schematic description. In order to recover the symmetric Lefschetz thimble $\mathcal{J}^{\mathrm{sym}.}_{r=1}$ in the limit $\ve\to +0$, we must sum $\mathcal{J}_{(1+\ve/2,\pi)}$ and $\mathcal{J}_{(1-\ve/2,0)}$ for the cancellation of the $\eta_2$ direction. 

What is the fate of each Lefschetz thimble $\mathcal{J}_{(1+\ve/2,\pi)}$, $\mathcal{J}_{(1-\ve/2,0)}$ in the limit $\ve \to +0$? 
Let us observe the behavior of integration on $\mathcal{J}_{(1-\ve/2,0)}$ in the limit $\ve\to +0$:
\bea
&&\int_{\mathcal{J}_{(1-\ve/2,0)}}\diff^2 z\exp\left[-\left(S_0+\ve\Delta S\right)/\hbar\right]
\nonumber\\
&\sim&-\im\int_0^{\infty}\lambda \diff \lambda \exp\left(-{(\lambda^2-1)^2\over 4\hbar}\right)\nonumber\\
&&\times\int_{-\infty}^{\infty}\diff \phi \exp\left(-{\ve\lambda\over \hbar} \cosh\phi\right). 
\eea
Here, we have set $\sigma_1=\lambda\cosh\phi$ and $\eta_2=\lambda \sinh\phi$. 
The integration of $\lambda$ is convergent, but the integration of $\phi$ is logarithmically divergent in the limit $\ve/\hbar\to +0$:
\be
\int\diff \phi \exp\left(-{\ve\lambda\over \hbar} \cosh\phi\right)\sim 2\ln {\hbar\over \ve}. 
\ee
This divergent integration is nothing but the integration along the critical orbit $r^2-(p_{\theta}/r)^2=1$. 
The exact same divergence with the opposite sign appears in the integration over $\mathcal{J}_{(1+\ve/2,\pi)}$, and we must sum them up to construct a convergent integration cycle in the limit $\ve\to0$. 
As already mentioned, this is identical with $\mathcal{J}^{\mathrm{sym.}}_{r=1}$ in the homological sense: 
\be
\mathcal{J}^{\mathrm{sym}.}_{r=1}=\mathcal{J}_{(1+\ve/2,\pi)}+\mathcal{J}_{(1-\ve/2,0)}. 
\ee
This might be surprising because the number of possible integration cycles changes under continuous change of parameters. Physically, this means that the number of solutions of the Dyson--Schwinger equation in the symmetric system is smaller than that of symmetry-broken systems. 

As we mentioned at the last paragraph of Sec.~\ref{sec:form}, we need to pick up half dimensions of the critical orbit $T^*S^{n-1}$ in order to construct the middle-dimensional integration cycle $\mathcal{J}^{\mathrm{sym.}}_{r=1}$. In the previous section, we picked them up in a specific way by rotating the Lefschetz thimble of the reduced system using the original symmetry $O(n)$. However, we did not explain why we cannot use imaginary directions of $O(n,\mathbb{C})$. 
The above analysis clearly shows the reason we have to give up imaginary directions of the complexified symmetry group: Otherwise, integration does not converge.

We have stuck to the case $\alpha=0^+$ in order to draw Figs.~\ref{fig:xyflow} and \ref{fig:xpyflow}, but our proposal is expected to be useful for generic $\alpha$. 
Since there is no Stokes jump in $0<\alpha<\pi$, Lefschetz thimbles at generic $\alpha$ are continuous deformations of Fig.~\ref{fig:schematic} (in $\mathbb{C}^2$). 
In order to span such cycles, flows should still go along $T^* S^1$ up to $\mathcal{O}(\ve)$ as shown in Fig.~\ref{fig:flow_critical_orbit}, and branch. 
The dynamics after the branch can be solved in a similar way to get Fig.~\ref{fig:rprflow}: $\ve\Delta H$ is expected to be negligible since critical conditions are broken at $\mathcal{O}(\ve^0)$. 

\subsection{\boldmath Symmetry breaking in $O(n)$ sigma models}\label{sec:symm_break_on}
Let us consider the case with general $n$. Then, the original action has $O(n)$ symmetry, and the perturbation $\Delta H$ breaks it into $O(n-1)$ symmetry:  
\be
O(n-1)=\left(\begin{array}{c|c}1&0\\\hline0&O(n-1)\end{array}\right)\subset O(n). 
\ee
In order to study the effect of $\Delta H$, we need nothing new. All the computations are reduced into the ones in the case of $O(2)$ symmetry as follows. 

Since the $O(n-1)$ symmetry is not broken at all, let us integrate it out. The $O(n-1)$-invariant variables are 
\be
\sigma_1,\; \sigma_2'=\sqrt{\sigma_2^2+\cdots+\sigma_n^2}. 
\ee
The original $O(n)$ invariance ensures that the unperturbed action can only depend on $\sigma_1^2+{\sigma'_2}^2$. 
Therefore, even after $O(n-1)$ symmetry is reduced, there still exists a continuous $O(2)$ symmetry. 
Thus, all of the previous argument for the system with explicit symmetry breaking holds. 
After computing downward flows in the reduced system, we can construct Lefschetz thimbles by rotating them under $O(n-1)$.

\section{Theory with fermions}\label{sec:fermion}

So far, we have assumed that the action $S$ is a polynomial of bosonic fields $\sigma_a$. 
Let us confirm that above properties of slow downward flows can also be observed in the system including fermions.  
This gives circumstantial evidence for universality of slow flows along critical orbits and of its relation to Lefschetz thimbles. 
Let $\psi_1,\psi_2$ be complex Grassmannian variables, and the interaction with fermions is introduced as 
\bea
&&\int \diff^2\psi\diff^2\psi^* \exp\mu(\psi_1^*\psi_1-\psi_2^*\psi_2)\nonumber\\
&&\times\exp\left[(\sigma_1+\im\sigma_2)\psi_1^*\psi_2^*+(\sigma_1-\im \sigma_2)\psi_2\psi_1\right]. 
\eea
This interaction looks similar to that of the BCS effective theory of superfluidity, by regarding $\sigma_i$ as effective fields for Cooper pairs and $\psi_i$ as elementary fermions. 
In this interpretation, the parameter $\mu$ corresponds to the chemical potential for fermion spin, which induces spin imbalance. 
The action has the following $O(2)$ symmetry, 
\bea
&\delta \sigma_1=-\sigma_2,\; \delta \sigma_2=\sigma_1,\nonumber\\ 
&\delta \psi_a={\im\over 2}\psi_a,\;
\delta \psi^*_a=-{\im\over 2}\psi^*_a. 
\eea
The Grassmannian integral gives the effective bosonic action,
\be
S_{\mathrm{eff}}[\sigma_a]=-\hbar\ln(\sigma_1^2+\sigma_2^2-\mu^2). 
\ee
Although this term has an ambiguity by a multiple of $2\pi \im$ due to logarithmic singularities, choice of branch does not affect the path integral because $e^{2\pi \im}=1$. 
This term produces a complex phase when $\mu\not=0$, and then it can cause the sign problem in Monte Carlo integration. 
One can also consider QCD-like models as fermionic systems with similar continuous symmetries, and detailed studies of them will be reported elsewhere \cite{Kanazawa:2014qma}. 

\begin{figure}[t]
\centering
\begin{minipage}{.5\textwidth}
\subfloat[$\mu<\sqrt{2\hbar}$]{
\includegraphics[scale=0.3]{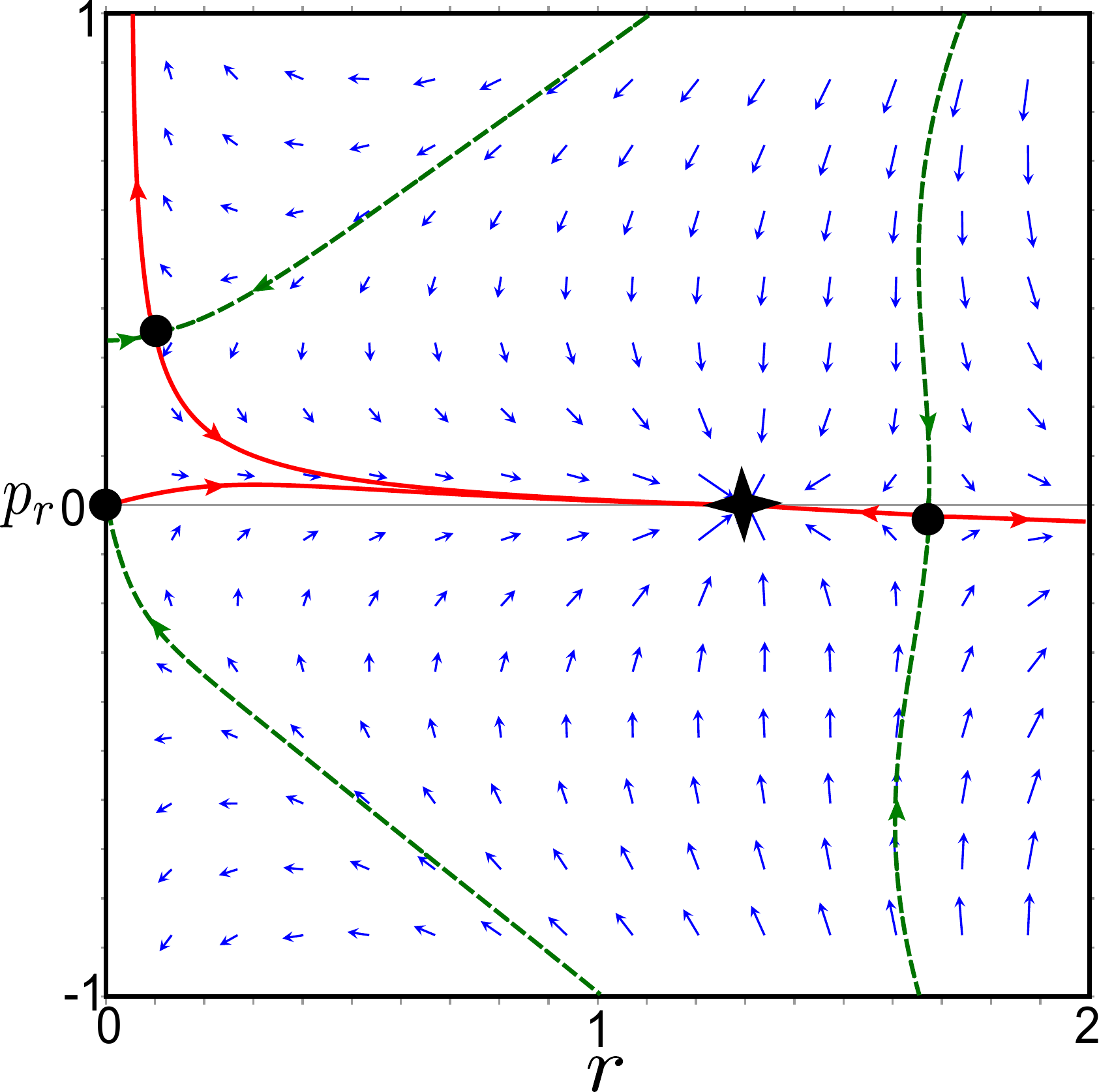}
}\end{minipage}
\begin{minipage}{.5\textwidth}
\subfloat[$\mu>\sqrt{2\hbar}$]{
\includegraphics[scale=0.3]{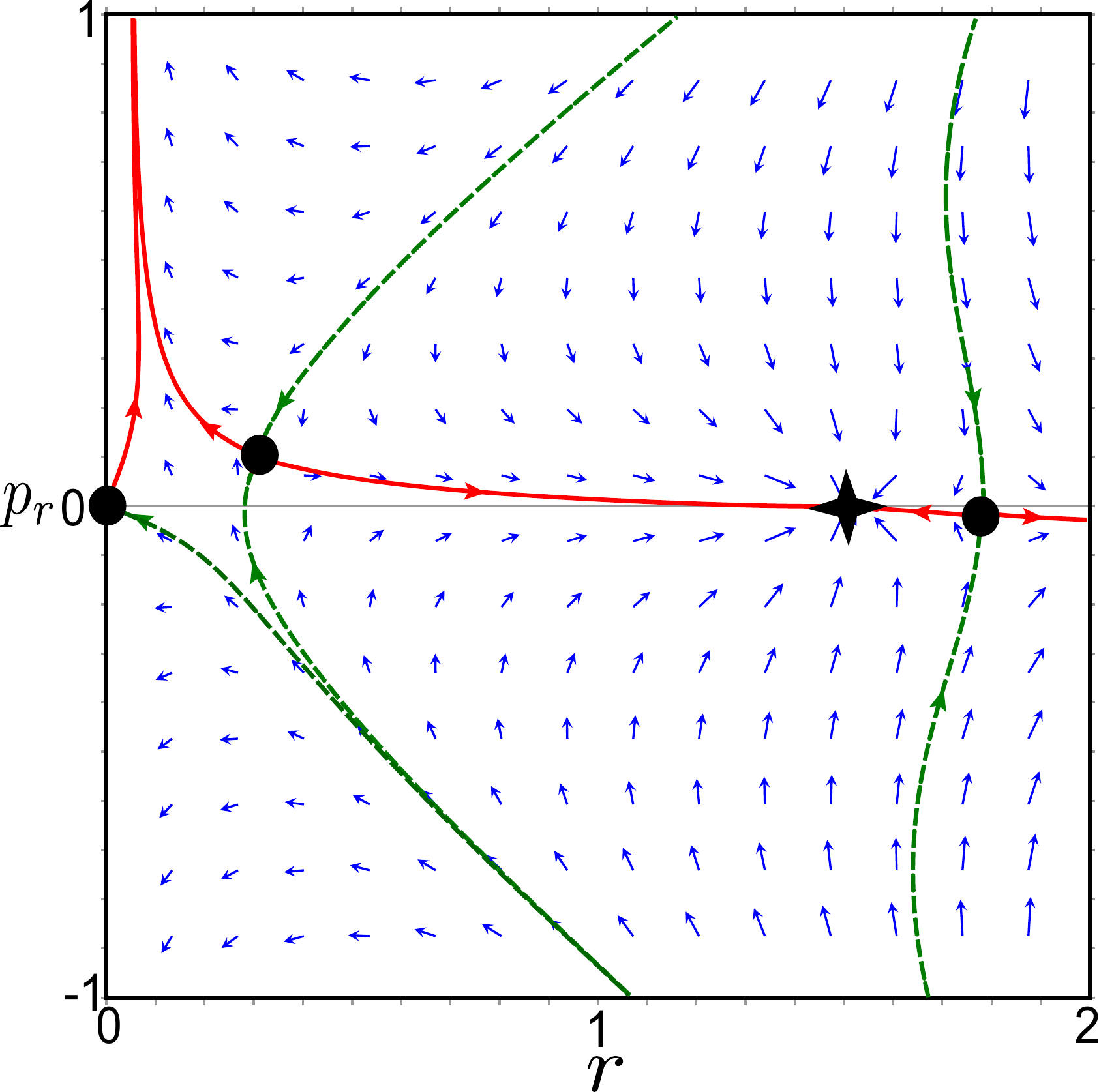}
}\end{minipage}
\caption{Behaviors of the downward flow equation for the model including fermions ($\mu=1.3$ and $1.5$, respectively, at $\alpha=0.1$, $\hbar=1$, and $p_{\theta}=0$). Red solid and green dashed curves represent Lefschetz thimbles and their homological duals of the reduced system, respectively. Blue arrows show Hamiltonian vector fields. }
\label{fig:rprflow_fermion}
\end{figure}

We consider structures of the Lefschetz thimble for the theory $S_{0+\mathrm{eff}}=S_0+S_{\mathrm{eff}}$. 
This change adds a new interaction term to the Hamiltonian: 
\bea
H_{\mathrm{eff}}&=&\mathrm{Im}S_{\mathrm{eff}}[\xi_a]\nonumber\\
&=&\hbar\tan^{-1}{2r p_r\over (p_r^2+p_{\theta}^2/r^2)-r^2+\mu^2}. 
\eea
The new term $H_{\mathrm{eff}}$ is not globally well defined in $\mathbb{C}^2\setminus\{\bm{\xi}^2=\mu^2\}$,  because of an ambiguity by a multiple of $2\pi\hbar$ associated with the choice of branch cuts. 
However, the flow equation is still well defined because it only contains derivatives of $H_{\mathrm{eff}}$, in which the ambiguity disappears. 
This observation is important for validity of the Picard--Lefschetz integration methods for fermionic systems (further details will be discussed in \cite{Kanazawa:2014qma}). 

The critical condition for the $O(2)$-symmetric system $S_{0+\mathrm{eff}}$ is given by 
\be
{\xi_a}\left((\bm{\xi}^2-1)-{2\hbar e^{-\im \alpha}\over \bm{\xi}^2-\mu^2}\right)=0, 
\ee
which can be solved as $\bm{\xi}=0$ or 
\be
\bm{\xi}^2=\xi_{\pm}^2\equiv{(\mu^2+1)\pm\sqrt{(\mu^2-1)^2+8\hbar e^{-\im\alpha}}\over 2}. 
\ee
At $\alpha=0^+$, $\xi_-$ becomes purely imaginary and $\xi_+$ is real if $\mu^2<2\hbar$, but both ones, $\xi_{\pm}$, remain real if $\mu^2>2\hbar$. 
This affects the structure of Lefschetz thimbles and their homological duals, and the Stokes jumping occurs in crossing $\mu=\mu_*\equiv\sqrt{2\hbar}$ when $\alpha=0^+$.  
Figure~\ref{fig:rprflow_fermion} shows behaviors of Lefschetz thimbles in the $r$$p_r$ plane at $p_{\theta}=0$ for $\mu=1.3$ and $\mu=1.5$ when $\alpha=0.1$ and $\hbar=1$.

\begin{figure}[t]
\centering
\begin{minipage}{.5\textwidth}
\subfloat[$\mu<\sqrt{2\hbar}$]{
\includegraphics[scale=0.3]{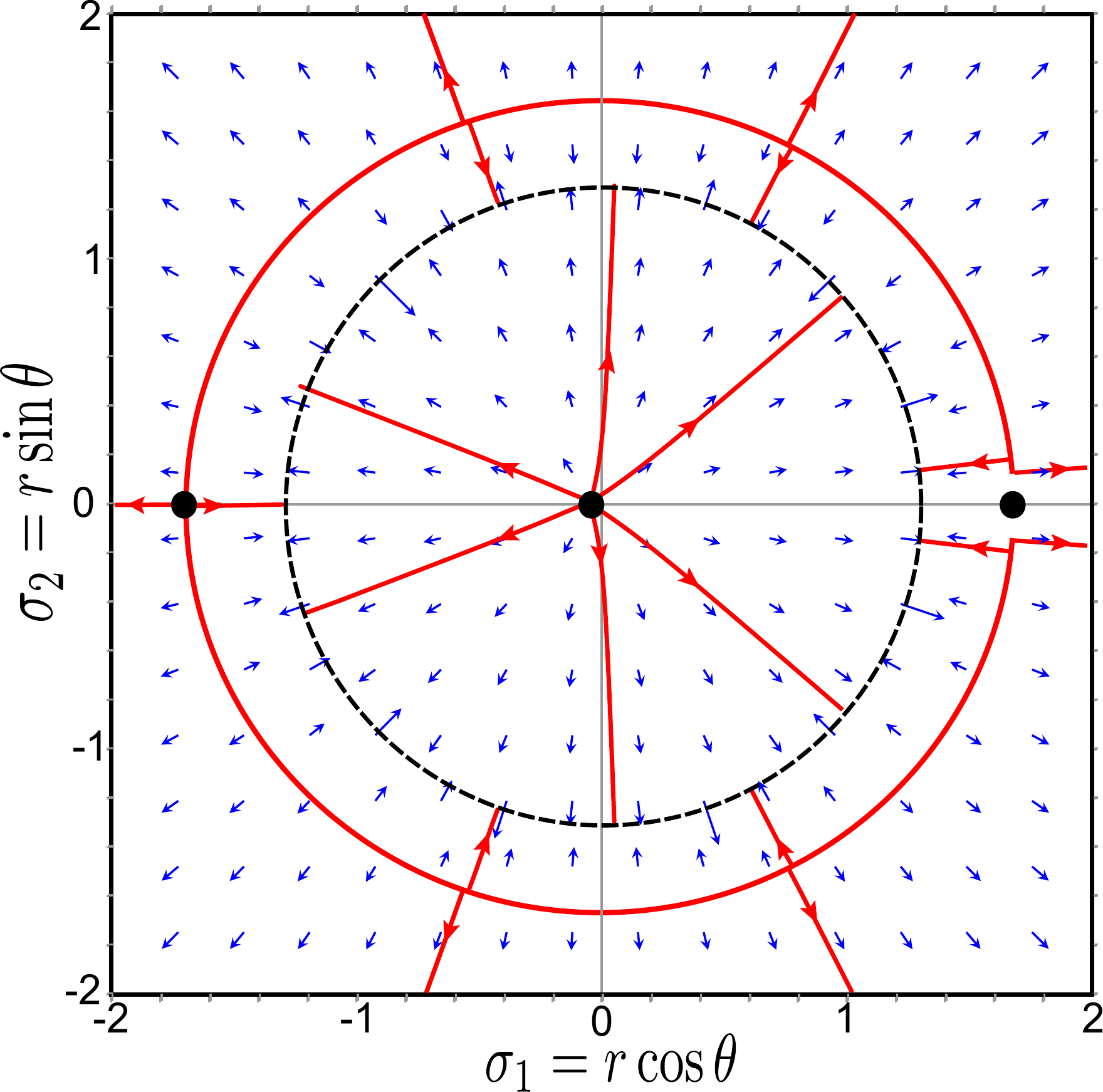}}
\end{minipage}
\begin{minipage}{.5\textwidth}
\subfloat[$\mu>\sqrt{2\hbar}$]{
\includegraphics[scale=0.3]{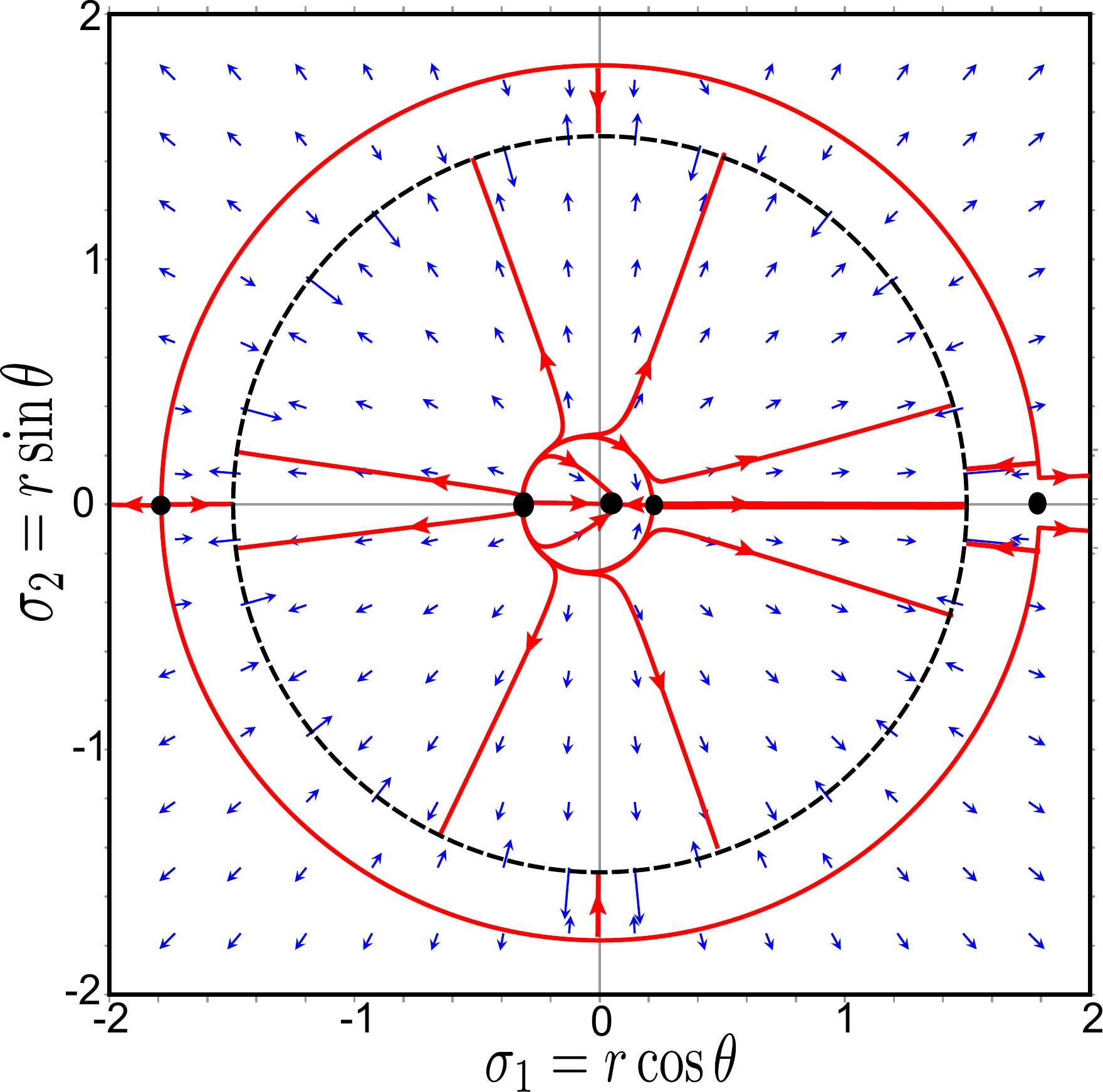}}
\end{minipage}
\caption{Behaviors of the downward flow equation (\ref{eq:hamilton_fermion_couple_simple}) for $\mu=1.3$ and $1.5$, respectively, at $\alpha=0$, $\hbar=1$, and $p_r=p_{\theta}=0$. 
The black dashed circle shows the set of logarithmic singular points $r=\mu$. We set $\ve=0.01$ in these figures. }
\label{fig:xyflow_fermion}
\end{figure}

Let us explicitly compute global structures of Lefschetz thimbles with the symmetry-breaking term $\ve \Delta H$ in the limit $\alpha=0^+$. 
Again, the condition $p_r=p_{\theta}=0$ gives a set of special solutions, which is important for our purpose to draw slow downward flows along pseudocritical orbits. 
The Hamilton equations for $r$ and $\theta$ in this condition are given as 
\begin{subequations}\label{eq:hamilton_fermion_couple_simple}
\begin{align}
&{\diff r\over\diff t}=r(r^2-1)-{2\hbar r\over r^2-\mu^2}+\ve\cos\theta, 
\label{eq:hamilton_fermion_simple01}\\
&{\diff \theta\over \diff t}=-{\ve}{\sin\theta\over r}. 
\label{eq:hamilton_fermion_simple02}
\end{align}
\end{subequations} 
The second term on the r.h.s. of (\ref{eq:hamilton_fermion_simple01}) is the effect of the additional interaction $H_{\mathrm{eff}}$, and the terms proportional to $\ve$ are coming from a symmetry-breaking perturbation $\ve \Delta H$ in (\ref{eq:break03}).

By adding the symmetry-breaking term (\ref{eq:break03}), the number of (semistable) critical points becomes five ($\xi=(0,0),(\pm \xi_{+},0), (\pm \xi_{-},0)$ up to $\mathcal{O}(\ve^0)$). 
If $\ve$ is assumed to be sufficiently small, only three of them, $(0,0)$ and $(\pm\xi_+,0)$, can contribute to the original path integral when $\mu<\sqrt{2\hbar}$, and all the five critical points do when $\mu>\sqrt{2\hbar}$. 
Figures~\ref{fig:xyflow_fermion}~(a) and (b) show typical behaviors of the downward flow (\ref{eq:hamilton_fermion_couple_simple})  at $p_{r}=p_{\theta}=0$ for $\mu<\sqrt{2\hbar}$ and for $\mu>\sqrt{2\hbar}$, respectively, at $\alpha=0^+$. 
We can again observe that downward flows travel slowly along critical orbits and branch into radial directions in order to span two-dimensional integration cycles. 

Figure~\ref{fig:xyflow_fermion} corresponds to Fig.~\ref{fig:xyflow} in the bosonic example, and behaviors of the slow motion along critical orbits turn out to be the same. 
Therefore, all of the previous arguments in the bosonic example can also be applied to this fermionic system, and we can readily obtain global structures of Lefschetz thimbles in the same manner. 


\section{Summary}\label{sec:summary}

The Picard--Lefschetz integration method with continuous symmetries is studied in detail through zero-dimensional $O(n)$ sigma models. 
We solve downward flow equations with small symmetry-breaking terms, and global structures of Lefschetz thimbles are identified. 
Moreover, we propose an efficient way to compute Lefschetz thimbles with approximate symmetry by extracting a general feature of downward flows in this model study. 

The explicit breaking term solves degeneracies of critical orbits, and thus the usual formulation of Lefschetz-thimble path integral becomes valid, in principle. 
However, approximate symmetries obstruct naive computation of downward flows: Downward flows first travel slowly along a pseudocritical orbit, and then branch into other directions. 
Tuning of the initial condition, which is a necessary step  in order to draw a set of flow lines for Lefschetz thimbles,  is thus very difficult. 
To circumvent this problem, we propose that downward flows on the pseudocritical orbit be solved at first. 
By starting and reverting a flow from a point in the vicinity of that downward flow, we can easily compute a downward flow branching at that point: This does not require severe tuning of initial conditions of the flow equation.

We believe that the property of downward flows observed in this model study is universal for flow equations with approximate symmetries. Once the critical conditions are broken at zeroth order, small perturbations become negligible and flows along symmetric direction cannot occur. 
Therefore, slow downward flows along the pseudocritical orbit are an important ingredient for creating middle-dimensional cycles. 
As a nontrivial check of this argument,  a fermionic system with approximate $O(2)$ symmetry is studied by using the same method, and slow motion along critical orbits can be observed in the exact same manner.

We also prove that only special combinations can survive as convergent integration cycles in the limit of symmetry restoration, and that these cycles correctly provide Lefschetz thimbles with the symmetry discussed in Ref.~\cite{Witten:2010cx}. 
This drastic change of possible integration cycles happens because integration along imaginary directions of critical orbits diverges as symmetry restores. 
We can now have a good understanding for the relation of Lefschetz thimbles with and without symmetry. 

It is still nontrivial how these properties play a role in the symmetry-breaking phenomenon. 
Thus, it is an interesting and important future study to describe spontaneous symmetry breaking of statistical physics from the viewpoint of the Picard--Lefschetz integration method. 
For that purpose, it would be important to scrutinize behaviors of Lefschetz thimbles in the thermodynamic limit. 

\begin{acknowledgments}
Y.~T. thanks Takuya Kanazawa for many useful discussions and comments on the manuscript. 
During writing this paper, the author had a chance to visit North Carolina State University, and he appreciates discussing many related stimulus topics on Lefschetz thimbles with Mithat {\"U}nsal, Aleksey Cherman, and Alireza Behtash.  
Y.~T. is supported by Grants-in-Aid for the fellowship of Japan Society for the Promotion of Science (JSPS) (No.25-6615). 
This work was partially supported by the RIKEN interdisciplinary Theoretical Science (iTHES) project, and by the Program for Leading Graduate Schools of Ministry of Education, Culture, Sports, Science, and Technology (MEXT), Japan.
\end{acknowledgments}

\appendix
\section{Flow equations at generic $\alpha$}\label{app:flow_generic}
In this appendix, we give a complete expression of the downward flow equation (\ref{Eq:coupled}) for generic $\alpha$. 
\begin{subequations}
\begin{align}
&{\diff r \over \diff t}=\cos\alpha\left\{r\left(r^2-{p_{\theta}^2\over r^2}-1\right)-3r p_r^2+\ve\cos\theta\right\}\nonumber\\
&\qquad+\sin\alpha\; p_r\left(1-3r^2+p_r^2+{p_{\theta}^2\over r^2}\right),
\label{eq:general_r}
\end{align}
\begin{align}
&{\diff p_r \over \diff t}=\cos\alpha\left\{p_r\left(1-3r^2 + p_r^2 -{ p_{\theta}^2\over r^2}\right)-\ve{p_{\theta}\over r}\sin\theta\right\}\nonumber\\
&
-{\sin\alpha}\left\{{r^4+p_{\theta}^2\over r^3}\left(r^2-p_r^2-{p_{\theta}^2\over r^2}-1\right)-2rp_r^2+\ve\cos\theta\right\},
\label{eq:general_pr}
\end{align}
\begin{align}
&{\diff \theta\over \diff t}=-\cos\alpha{\ve\sin\theta+2p_r p_{\theta}\over r}\nonumber\\&\qquad
-\sin\alpha{p_\theta\over r^2}\left(r^2-p_r^2-{p_{\theta}^2\over r^2}-1\right),
\label{eq:general_theta}
\end{align}
\begin{align}
\label{eq:general_ptheta}
&{\diff p_{\theta}\over \diff t}
=\ve\cos \alpha\left(p_r\sin\theta+{p_{\theta}\over r}\cos\theta\right)+\ve \sin\alpha\; r \sin\theta. 
\end{align}
\end{subequations}
Since we have set $\alpha=0^+$, terms proportional to $\sin\alpha$ do not have any role for the most part of downward flows. Only when the flow emanating from one critical point comes into the vicinity of another critical point, does that term designate the way of the flow to circumvent the another critical point.

\bibliography{lefschetz,./ref}

\end{document}